# What retards the response of graphene based gaseous sensor

Hui-Fen Zhang[a], Bo-Yuan Ning[b], Tsu-Chien Weng[b], Dong-Ping Wu[c,*], Xi-Jing Ning[a,*]

[a]*Applied Ion Beam Physics Laboratory, Institute of Modern Physics, Fudan University, Shanghai 200433, China*

[b]*Center for High Pressure Science & Technology Advanced Research, Shanghai, 202103, China*

[c]*State Key Laboratory of ASIC & System, Fudan University, Shanghai 200433*

**Abstract** Graphene based sensor to gas molecules should be ultrasensitive and ultrafast because of the single-atomic thickness of graphene, while the response is not fast. Usually, the measured response time for many molecules, such as CO, $NH_3$, $SO_2$, $CO_2$ and $NO_2$ and so on, is on the scale of minutes or longer. In the present work, we found via *ab initio* calculations there exists a potential barrier larger than 0.7 eV that hinders the gas molecule to land directly at the defective sites of graphene and retards the response. An efficient approach to the problem is suggested as modifying the graphene sheet with other molecules to reduce the potential barrier and was demonstrated by a graphene sheet modified by $Fe_2O_3$ molecules that shows fast response to $H_2S$ molecule, and the calculated response time is close to the measured one, 500 μs.

**Keywords**: graphene; gaseous sensor; diffusion model; statistical physics; *ab initio* calculations; surface

## 1. Introduction

The discovery of graphene has opened unprecedented opportunity that promises ultrasensitive and ultrafast electronic sensors due to its high surface-volume ratio, low electronic noise and exceptional transport properties associated with its unique high crystalline single-atom thick two-dimensional structure [1]. Indeed, the ultra-sensitivity has been proved by the graphene based gaseous sensor (GBGS) that works by measuring the conductance changing induced by the molecules adsorbed on the graphene sheet [2-6]. However, the measured response rate is not fast, i.e., the response time disperses in a large range from tens of seconds to thousands of minutes [7]. To see what retards the response rate and how to overcome the blocks need an extensive understanding of the working mechanism of GBGS on atomistic level.

Very recently, we extensively calculated the rate of adsorption and desorption of gas molecules on graphene surface and found that the balance between the adsorption and desorption would be reached within several microseconds under common experimental conditions [8], implying that the response time should be on the scale of microseconds if the molecules adsorbed in the perfect region, which occupies most area of a graphene sheet, could induce the conductance change of graphene. Based on this fact, the mechanism for GBGS working was suggested as that the gas molecules landing on the perfect region do not contribute to the conductance, which is in agreement with previous theoretical results [9], and have to undergo diffusion process to arrive at the defective sites, where they affect the conductance of graphene. That is, the diffusion process is responsible for the slow response of GBGS. There exists a query, however, why the gas molecules do not land directly at the defective sites inducing changes of the conductance immediately?

In the present work, we firstly examined if there exists a potential barrier hindering the gas molecules to land directly at the defective sites of the graphene sheet, and found by *ab initio* calculations it is true for CO, $CO_2$, $NH_3$, and $SO_2$ molecules, to which the observed response of GBGS is slow [4, 10-15]. Then we explored the possibility to remove or reduce the potential barrier by modifying the surface of graphene sheet with other molecules. As an example, we simulated the process of a $H_2S$ molecule approaching to a $Fe_2O_3$ molecule adsorbed on a graphene sheet, and show that a barrier of only 0.14 eV or smaller exists and therefore the conductance response should be fast, which is in agreement with the previous observations [16]. Finally, it was predicted that the response of GBGS to the NO gas molecules would be very fast (the response time is on the scale of about 0.01s) if the concentration of the gas molecules is higher than 100 ppm.

## 2. Theoretical method

On the atomistic level, the defective region of a graphene sheet covered by gas molecules depends not only on a potential barrier $E_a$, met by a gas molecule approaching to the defective region, but also on a barrier (desorption barrier), $E_d$, met by a molecule escaping from the defective region [Fig. 1], as well as on the gas concentration $n$ and so on. According to our kinetic model for the adsorption and desorption of gas molecules on a surface [8], the fractional coverage $\theta$ of a defective region of a graphene sheet is determined by

$$\frac{d\theta}{dt} =$$

$$n\sqrt{\frac{k_B T_g}{2\pi m}} \cdot \frac{1}{Z_a} \int_{E_a}^{E_b} \sqrt{\varepsilon}\, e^{-\frac{\varepsilon}{k_B T_g}} d\varepsilon \cdot S \cdot (1-\theta) - \frac{\frac{1}{Z_d}(\int_{E_d}^{\infty} \sqrt{\varepsilon}\, e^{-\frac{\varepsilon}{k_B T_d}} d\varepsilon)^2}{\int_{E_d}^{\infty} (\delta t)\sqrt{\varepsilon}\, e^{-\frac{\varepsilon}{k_B T_d}} d\varepsilon} \cdot \theta, \quad (1)$$

where $Z_a = \int_0^{\infty} \sqrt{\varepsilon}\, e^{-\frac{\varepsilon}{k_B T_g}} d\varepsilon$, $Z_d = \int_0^{\infty} \sqrt{\varepsilon}\, e^{-\frac{\varepsilon}{k_B T_d}} d\varepsilon$ with $k_B$ the Boltzmann constant and $\delta t = \sqrt{m} \int_{z_0}^{z_0+d} dz/\sqrt{2(\varepsilon - V(z))}$ with $V(z)$ felt by a molecule of mass $m$ located at $z_0$ moving along the $z$ direction by a distance $d$ to escape from the potential valley [Fig. 1]. In addition, $S$ is the effective area of a molecule on the surface, and $E_b$ is introduced for considering the fact that the molecules with kinetic energy larger than $E_b$ will immediately escape from the surface just after its colliding with the surface. The first term in the right hand of Eq. (1) is the adsorption rate of the gas molecules, while the second one is the desorption rate. According to the mechanism for GBGS working [8], the adsorption of gas molecules at defective sites instead of perfect sites is responsible for the conductance change of graphene, so the balance time (BT) between the adsorption and desorption of the gas molecules at defective sites determined by Eq. (1) with $d\theta/dt = 0$ should be on the same scale as the response time defined experimentally if the BT is shorter than the time need by the diffusion of gas molecules landing in the perfect regions to the defective sites. Usually, GBGS works at room temperature, so the gas temperature $T_g$ and desorption temperature $T_d$, i.e. the temperature of the substrate, are set as 300 K in our calculations.

For determining the $E_a$ and $E_d$, we fully relaxed a graphene sheet with a single vacancy (vG) at the center region firstly, and then put a gas molecule on (or above) the vacancy and moved it perpendicularly departing from (or approaching to) the vG step by step to obtain the dependence of the total energy on the distance between the gas molecule and the substrate. All the calculations were performed in the Vienna ab initio simulation package (VASP) with projected augmented-wave (PAW) potential employed to describe the electron-ion interaction and Perdew-Burke-Ernzerhof (PBE) of generalized gradient approximation (GGA) to consider the exchange-correlation interaction. We used a $3 \times 3$ supercell for vG with Brillouin zone k-mesh of $5 \times 5 \times 1$. The vacuum thickness, distance between two adjacent graphene layers is 20 Å in all of the calculations. During the optimization of all of the structures, the vacuum thickness is frozen but the other two vectors are fully relaxed. The kinetic energy cut-off is 400 eV. The electronic self-consistency will stop when the difference of the energy is smaller than $1\times10^{-6}$ eV and the force acted on each atom is less than 0.01 eV/Å.

In order to compare our calculated results with others, the binding energy $E_{bind}$ of a molecule on graphene is defined as

$$E_{bind} = E_{gra+molecule} - E_{gra} - E_{molecule} \qquad (2)$$

where $E_{gra+molecule}$ is the fully relaxed total energy of a graphene sheet adsorbed by a molecule. $E_{gra}$ and $E_{molecule}$ correspond to the energy of the graphene and an isolated molecule.

## 3. Results and discussions

### A. Adsorption on vG

The total energy ($E_T$) of the vG + CO ($NH_3$, $SO_2$, $CO_2$ or NO) system as function of the distance between the molecule and the sheet was obtained by moving the C atom of CO or $CO_2$, the N atom of $NH_3$ or NO, or the S atom of $SO_2$ molecule step by step (fixing the $z$- coordinate with the $x$- and $y$- coordinates free) approaching to (or departing from) the vacancy of the vG sheet fully relaxed. As shown in Fig. 2 for a CO molecule approaching to the vG, the $E_T$ displays a platform firstly and then increases suddenly up to a barrier of 3.89 eV, corresponding to binding of the C atom of the CO molecule with a C atom around the vacancy [inset

(a) of Fig. 2]. With further approaching to the vG, the $E_T$ gradually decreases until reaches a metastable state, corresponding to a configuration shown in inset (b), where the CO tilts with its C atom bonded by a C atom around the vacancy and the bond length is 1.33 Å, very close to the length of C=C double bond (1.34 Å). This result is in good agreement with the results of ref. [9, 17]. From this configuration, the CO molecule crosses another barrier of 0.18 eV and inserts its C atom into the vacancy with the O atom above the center of the near C−C bridge [inset (c)], which is similar to the optimized stable structure due to Nacir Tit et al. [18]. Fig. 2 shows clearly that a CO molecule approaching to the vG will meet an adsorption barrier of 3.89 eV ($E_a$ = 3.89 eV) to land on a metastable state and has to overcome another barrier of 0.18 eV to form the most stable configuration, from which the CO molecule must own a kinetic energy larger than 10.1 eV to escape ($E_d$ = 10.1 eV) from the vG.

According to Eq. (1) with $E_a$ = 3.89 eV, the adsorption rate of the CO molecule on the single vacancy with an area of about 10 Å$^2$ occupied by a molecule is zero even if the gas concentration $n$ is as large as 1 atm and $E_b$ set as infinity, indicating that the CO molecule cannot directly land on the vacancy of vG. So the molecules landing in the perfect region have to undergo a diffusion process to reach the defective sites, which retards the response of GBGS. As an example, the response time of GBGS to CO gas with a concentration of 100 ppm under room temperature was measured to be 15 min [4].

For a NH$_3$ molecule approaching to the vG, the $E_T$ jumps from a platform up to a barrier of 1.76 eV, corresponding to binding of the N atom with a C atom around the vacancy [inset (a) of Fig. 3], and the system finally form a stable structure [inset (b)], where a C atom drawn off the vG sheet by 1.03 Å binds with the N atom with a bond length of 1.458 Å, which is in good agreement with the result of ref. [19]. That is, there exists a barrier $E_a$ of 1.76 eV for the NH$_3$ molecule to land at the vacancy and the desorption barrier $E_d$ is 2.96 eV.

Based on Eq. (1) with $E_a$ = 1.76 eV, the adsorption rate of the NH$_3$ molecule on the vacancy is slower than $2.3 \times 10^{-22}$/s if the concentration of the gas is lower than 1 atm. So, the NH$_3$ molecule can hardly directly land on the vacancy of a graphene sheet and the long response time ( > 1 h) observed in experiments with the gas concentration of 58 ppm [10] should be attributed to the diffusion process of the molecule from the perfect regions to the defects where they induce the conductance change.

For a SO$_2$ molecule approaching to the vG [Fig. 4], the $E_T$ crosses a barrier of 0.75 eV from a platform, and then suddenly drops, which corresponds to binding of the S atom with two of the three C atoms around the vacancy [inset (a) of Fig. 4]. With further approaching to the sheet, the $E_T$ decreases until the molecule arrives at a stable state, where the S atom and an O atom bind with all the three C atoms around the vacancy [illustration (b)] with a binding energy of -2.26 eV, which is consistent with the results of ref. [20].

Based on the Eq. (1) with $E_a = 0.75$ eV, the adsorption rate for gas $SO_2$ with a concentration of 1 atm on the vacancy is $1.3 \times 10^{-9}$/s, i.e. covering the vacancy by a $SO_2$ molecule needs about 31.7 years. Indeed, in an experiment due to Ren et al. [13], the response time of GBGS to $SO_2$ gas with a concentration of 50 ppm is about 30 min, implying that the $SO_2$ molecules responsible for the conductance change at defective sites come from the perfect regions of vG via diffusion rather than directly land at the defective sites.

When a $CO_2$ molecule approaches to the vG, it crosses a barrier of 1.05 eV [Fig. 5] and then the $E_T$ displays a sudden drop to a valley, where the C and one of the O atoms of the $CO_2$ molecule bind with two C atoms around the vacancy [inset (a) of Fig. 5]. From this configuration, the molecule will meets another barrier of 0.45 eV to arrive at a stable state [inset (b)], which is similar to the result of ref. [21]. According to Eq. (1), covering the vacancy by a $CO_2$ molecule needs about $3.2 \times 10^6$ years, implying that the response of GBGS to $CO_2$ gas is slow because the molecules landing in the perfect region have to undergo a diffusion process to reach the vacancy. So the response rate depends on the roughness of a graphene sheet, which significantly affects the diffusion rate. Experimentally, the response time of $CO_2$ gas of GBGS with the graphene sheet by electrochemical exfoliated method is 11 s [14], which is slightly longer than the one (8 s) with the mechanical cleaved graphene [15] for the same gas concentration at room temperature.

Differently from the other molecules, a NO molecule could directly arrive at a valley [Fig. 6], where the N atom binds with one C atom around the vacancy [inset (a) of Fig. 6], which is same as the result of ref. [9]. In the realistic process of the molecule approaching to the vG, the kinetic energy of the molecule at this metastable site [inset (a)] transferred from the potential energy is as large as 2.5 eV, which is large enough for the molecule to cross a barrier of 0.8 eV to arrive at a deeper valley [Fig. 6], where the N atom is just above the missing C atom by 0.55 Å [inset (b)].

According to Eq. (1) with $E_a$ = 0 eV and $E_d$ = 5.84 eV, the BT of the adsorption and desorption on the vacancy is proportional to the gas concentration [inset (c) of Fig. 6]. When the concentration is 100 ppm, the balance time is about 0.01 s, corresponding to a very fast response. However, if the concentration decreases down to 200 ppt, the BT is as long as $10^4$ s. An experiment due to Chen et al. [1] shows that the response time of GBGS to NO gas with a concentration of 200 ppt is about 300 s, which is much shorter than the calculated time, $10^4$ s, implying that the NO molecules landing in the perfect region diffuse into the defective sites to induce the changes of the conductance.

**B. Adsorption on modified vG**

Based on the above results, a possible way to raise the response rate of GBGS is to reduce or remove the adsorption barrier that impedes the gas molecules landing directly at the defective sites of graphene. Stimulated by an experiment on the response of GBGS to $H_2S$ gas with the graphene modified by $Fe_2O_3$ molecules [16], we examined the potential barrier for a $H_2S$ molecule approaching to a $Fe_2O_3$ molecule adsorbed at the perfect graphene (pG) or a single vacancy of vG. Specifically, the structure of a $Fe_2O_3$ molecule on the graphene sheet was optimized firstly, and then a $H_2S$ molecule was placed above or beside the $Fe_2O_3$ molecule. After fully relaxing, the $z$- coordinate (perpendicular to the graphene sheet) of the S atom was changed step by step to depart from the graphene for calculating the $E_T$ as the function of $z$.

The most stable structure of a $Fe_2O_3$ molecule adsorbed on a perfect graphene sheet is shown in Fig .7a, where the flat $Fe_2O_3$ molecule is perpendicular to the surface with one of the Fe atoms just above the center of the six-ring, which is significantly different from the one of the $Fe_2O_3$ molecule curved at the vacancy of vG with one of the Fe atom bonded with the three C atoms around the vacancy [Fig .7b]. The binding energy (-5.997 eV) of the $Fe_2O_3$ molecule on the vacancy is much larger than the one (-1.25 eV) of the molecule on perfect graphene.

When a $H_2S$ molecule approaches to the $Fe_2O_3$ molecule at the perfect site, it crosses over a barrier of 0.14 eV into a potential valley [Fig. 8] with the configuration transferred from inset (b) to inset (a) of Fig. 8. When the molecule approaches to the $Fe_2O_3$ at the vacancy, however, it directly arrives at the most stable state without energy barrier [Fig. 9] and the corresponding configuration [the inset of Fig. 9] is obviously different from the one shown in inset (a) of Fig. 8.

Based on the Eq. (1), the BT for $H_2S$ molecules adsorbed by a $Fe_2O_3$ molecule on perfect graphene ($E_a$ = 0.14 eV, $E_d$ = 0.67 eV) is 15 μs, and the corresponding fractional coverage $\theta_s$ is 2.1 × 10$^{-6}$. If the $Fe_2O_3$ molecule locates at the vacancy of the vG, the BT reduces to 3.9 × 10$^{-9}$ s because $E_a$ = 0 eV and the corresponding $\theta_s$ decreases down to 3.7 × 10$^{-8}$ because the desorption energy ($E_d$ = 0.31 eV) is small. In the realistic process for modifying a graphene sheet with $Fe_2O_3$ molecules [16], the defects can trap the molecules, so do the perfect sites because the binding energy ($E_{bind}$) of the $Fe_2O_3$ molecule with the pG is -1.25 eV. Considering that the fractional coverage $\theta_s$ of $Fe_2O_3$ molecules at the vacancy by the $H_2S$ molecules is two orders smaller than the one for the $Fe_2O_3$ at the perfect sites, the response time of the GBGS should be on the scale of 15 μs, which approaches to the measured one, 500 μs [16].

## 4. Conclusion

Based on the mechanism for GBGS working, the response should be very fast if the gas molecules can directly land at the defective sites of graphene. However, our calculations via DFT show that a barrier of 0.75 ~

3.9 eV exists for CO, NH$_3$, SO$_2$, CO$_2$ and NO$_2$ molecule approaching to the defective sites. It is the barrier that retards the response of GBGS, and reducing or removing the barrier is the way to raise the response rate. In addition, we predicted that the response time of GBGS to gas NO with concentrations higher than 100 ppm is on the scale of 0.01s, which might be confirmed by future experiments.

**Acknowledgement**

This work was supported by the National Natural Science Foundation of China (Nos. 21727801, 61474028, 61774042); the Shanghai Municipal Natural Science Foundation (No. 17ZR1446500); and the National S&T Project 02 (No. 2013ZX02303-004) and "First-Class Construction" project of Fudan University.

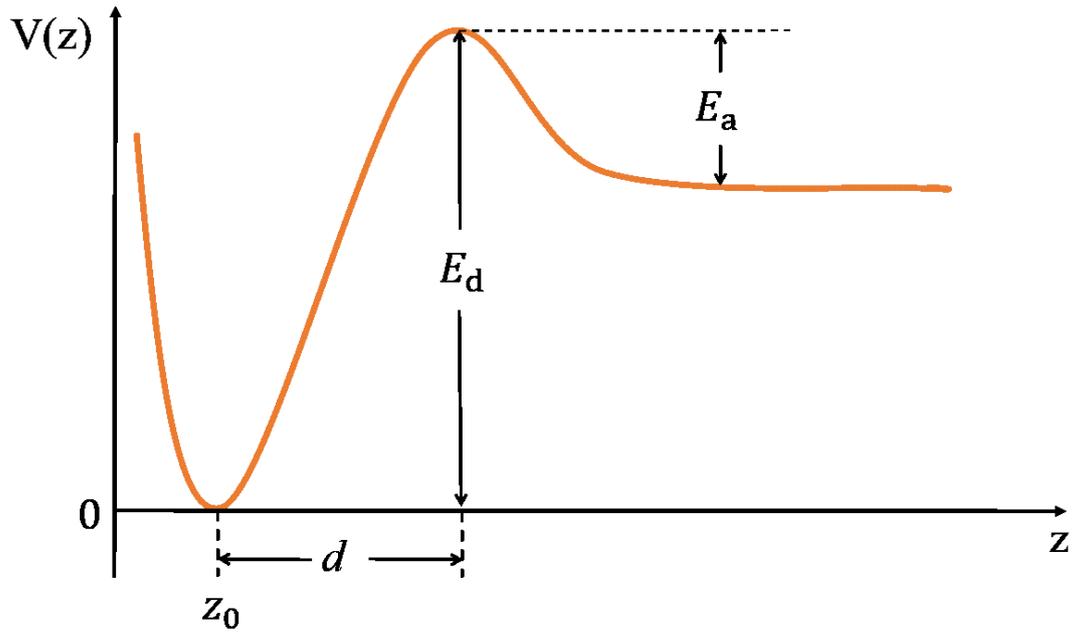

Fig. 1: Schematic of the potential felt by a molecule moving along the z-axis perpendicular to the surface of graphene

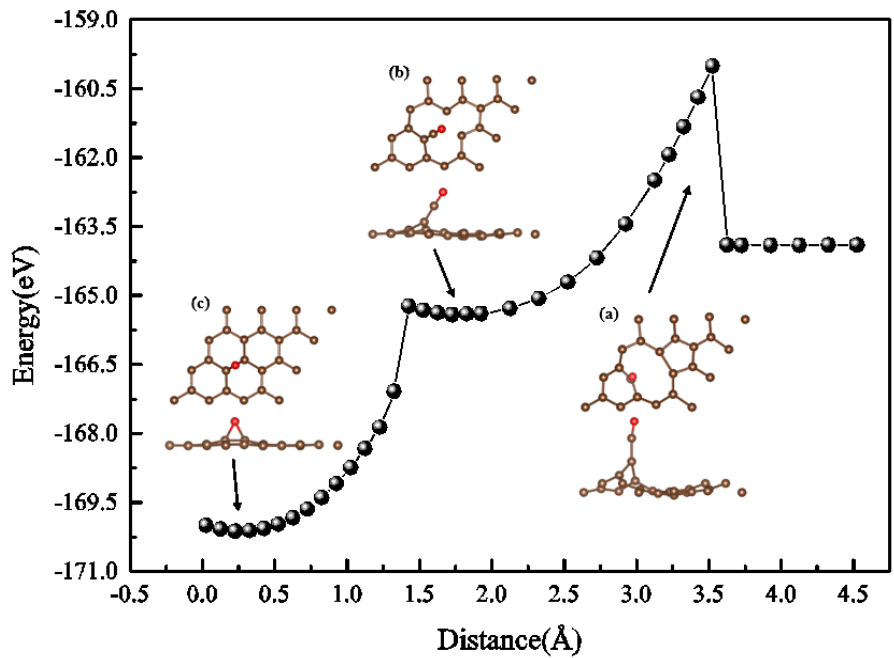

Fig. 2: The dependence of the $E_T$ of vG + CO system on the distance between the C atom and surface of vG with corresponding configurations of the system, inset (a), (b) and (c), where the dark wine and red ball represent the C and O atom, respectively.

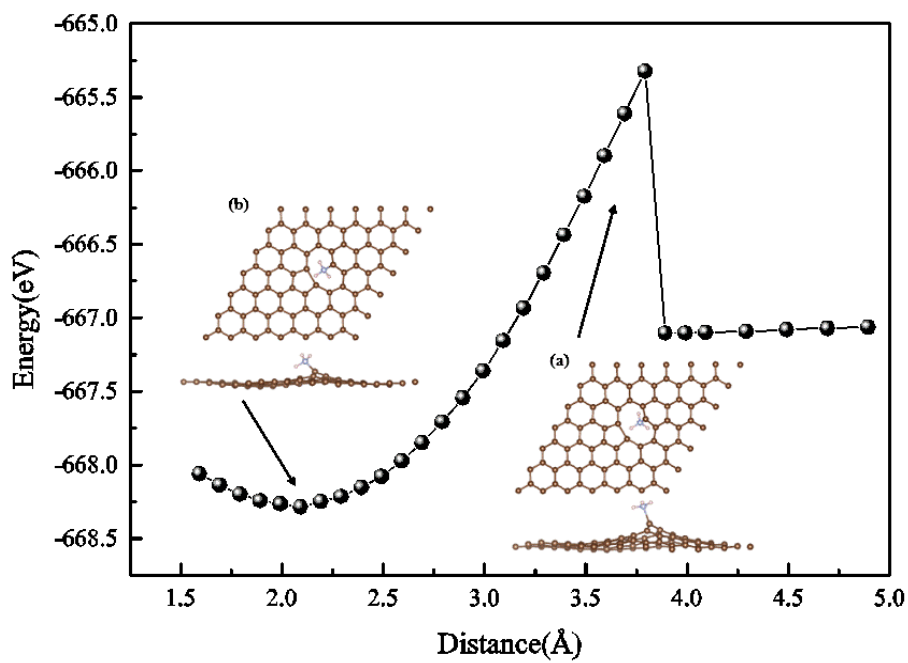

Fig. 3: The dependence of the $E_T$ of vG + NH$_3$ system on the distance between the N atom and the surface of vG with corresponding configurations of the system, inset (a), (b), where the dark wine, light blue and light pink ball represent the C, N and H atom, respectively.

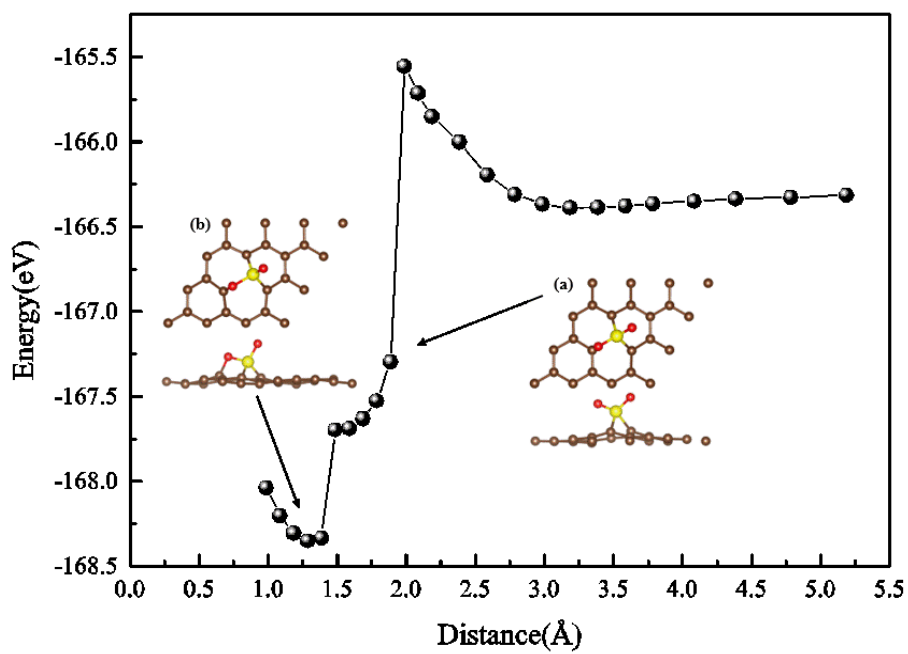

Fig. 4: The dependence of the $E_T$ of vG + $SO_2$ system on the distance between the S atom and the surface of vG with corresponding configurations of the system, inset (a), (b), where the dark wine, yellow and red ball represent the C, S and O atom, respectively.

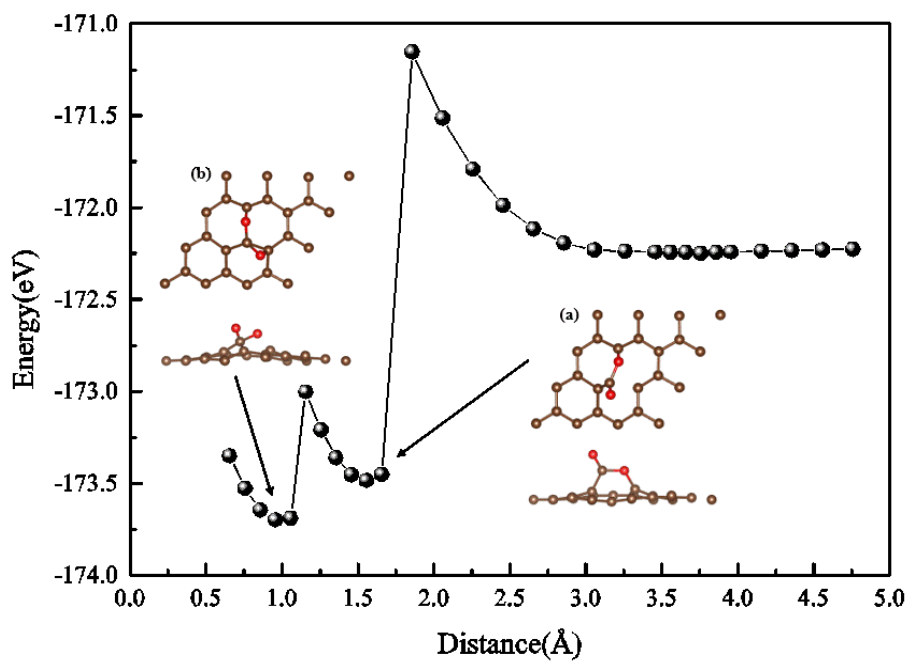

Fig. 5: The dependence of the $E_T$ of vG + $CO_2$ system on the distance between the C atom and the surface of vG with corresponding configurations of the system, inset (a), (b), where the dark wine and red ball represent the C and O atom, respectively.

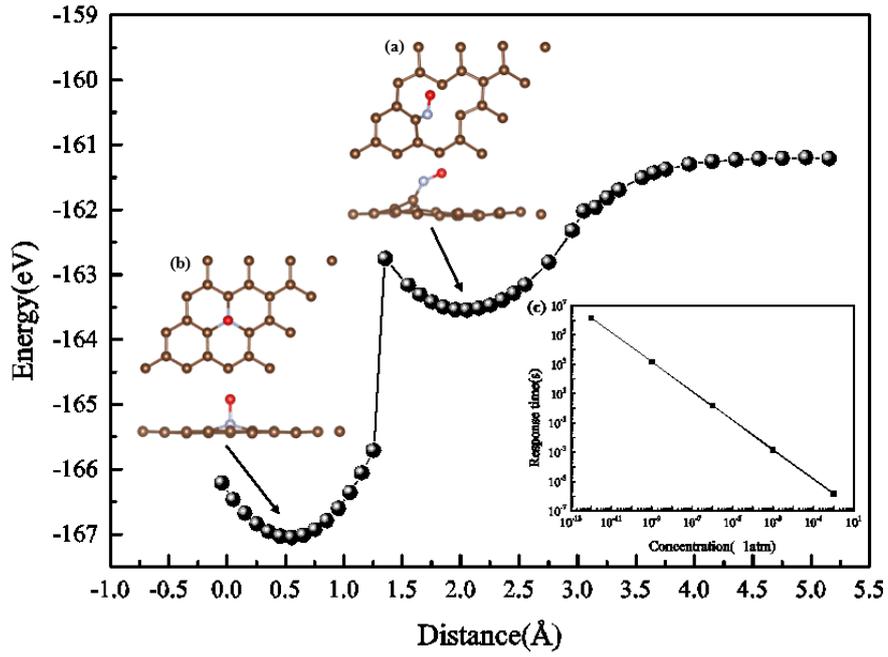

Fig. 6: The dependence of the $E_T$ of vG + NO system on the distance between the N atom and the surface of vG with corresponding configurations of the system, inset (a), (b), where the dark wine, light blue and red ball represent the C, N and O atom, respectively.

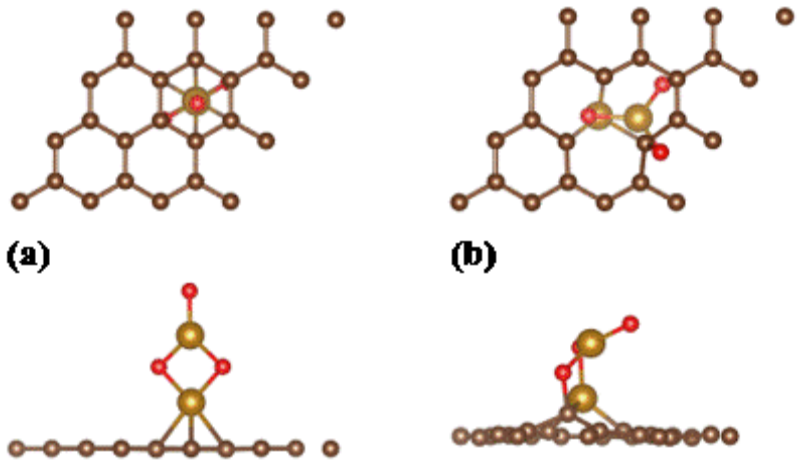

Fig. 7: The optimized configuration of a $Fe_2O_3$ molecule adsorbed on perfect graphene sheet (a) or at a vacancy of the vG (b), where the dark wine, light wine and red ball represent the C, Fe and O atom, respectively.

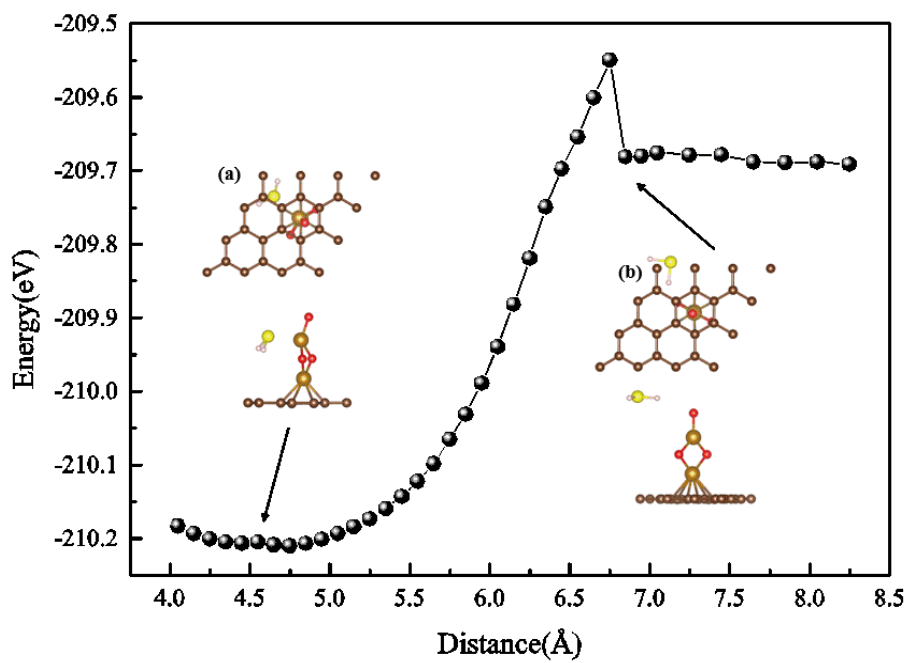

Fig. 8: The dependence of the $E_T$ of pG + $Fe_2O_3$ + $H_2S$ system on the distance between the S atom and the surface of vG with corresponding configurations of the system, inset (a), (b), where the dark wine, light wine, light pink, yellow and red ball represent the C, Fe, H, S and O atom, respectively.

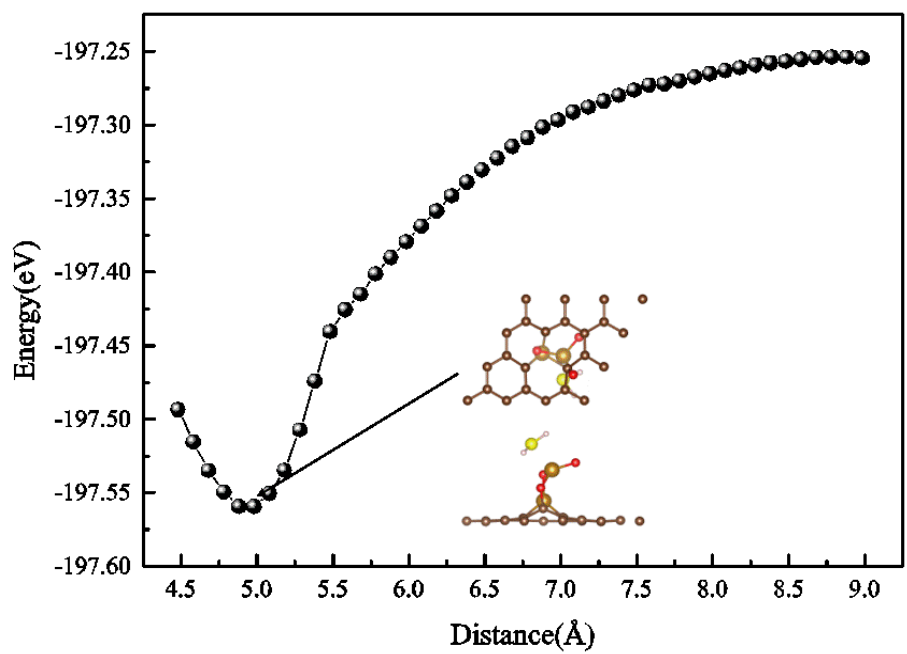

Fig. 9: The dependence of the $E_T$ of vG + $Fe_2O_3$ + $H_2S$ system on the distance between the S atom and the surface of vG with corresponding configurations of the system, inset, where the dark wine, light wine, light pink, yellow and red ball represent the C, Fe, H, S and O atom, respectively.